\documentclass{iaus}
\usepackage{graphicx}
\DeclareGraphicsExtensions{.eps, .ps}

\title[Light Neutron-Capture Elements in PNe] 
{The Abundances of Light Neutron-Capture Elements in Planetary Nebulae}

\author[N.\ C.\ Sterling \& H.\ L.\ Dinerstein]   
{N.\ C.\ Sterling$^1$
~\and Harriet L.\ Dinerstein$^1$}

\affiliation{$^1$University of Texas, Department of Astronomy, 1~University Station, C1400, Austin, TX, 78712-0259 \break
  email: sterling@astro.as.utexas.edu, harriet@astro.as.utexas.edu}

\pubyear{2006}
\volume{234}  
\pagerange{0--0}
\date{5-3-06}
\jname{Planetary Nebulae in our Galaxy and Beyond}
\editors{M. J. Barlow \& R. H. M\'endez, eds.}
\begin{document}

\maketitle

\begin{abstract}

We present preliminary results from a large-scale survey of the neutron(\emph{n})-capture elements Se and Kr in Galactic planetary nebulae (PNe).  These elements may be produced in PN progenitors by \emph{s}-process nucleosynthesis, and brought to the stellar envelope by third dredge-up (TDU).  We have searched for [Kr~III]~2.199 and [Se~IV]~2.287~$\mu$m in 120 PNe, and detected one or both lines in 79 objects, for a detection rate of 66\%.  In order to determine abundances of Se and Kr, we have added these elements to the atomic database of the photoionization code CLOUDY, and constructed a large grid of models to derive corrections for unobserved ionization stages.  Se and Kr are enriched in $\sim$73\% of the PNe in which they have been detected, and exhibit a wide range of abundances, from roughly solar to enriched by a factor of 10 or more.  These enrichments are interpreted as evidence for the operation of the \emph{s}-process and TDU in the progenitor stars.  In line with theoretical expectations, Kr is more strongly enhanced than Se, and the abundances of both elements are correlated with the carbon abundance.  Kr and Se are strongly enhanced in Type~I PNe, which may be evidence for the operation of the $^{22}$Ne neutron source in intermediate-mass AGB stars.  These results constitute the first broad characterization of \emph{s}-process enrichments in PNe as a population, and reveal the impact of low- and intermediate-mass stars on the chemical evolution of trans-iron elements in the Galaxy.

\end{abstract}

\firstsection 
\section{Introduction}

Low- and intermediate-mass stars (1--8~M$_{\odot}$), the progenitors of planetary nebulae (PNe), produce roughly half of the isotopes of neutron(\emph{n})-capture elements (atomic number $Z>30$) in the Universe.  During the thermally-pulsing asymptotic giant branch (AGB) phase, these stars create \emph{n}-capture elements via slow \emph{n}-capture nucleosynthesis, or the ``\emph{s}-process.''  Neutrons are released in the intershell region primarily through $^{13}$C($\alpha$,\emph{n})$^{16}$O, and to a lesser extent $^{22}$Ne($\alpha$,\emph{n})$^{25}$Mg.  Iron-peak seed nuclei capture the neutrons during the interpulse phase, and are transformed into heavier nuclei by a series of $\beta$-decays and \emph{n}-captures.  The intershell material, enriched in \emph{n}-capture elements and C, is conveyed to the stellar envelope by the third dredge-up (TDU) in stars with initial masses above 1.5~M$_{\odot}$, and is dispersed into the surrounding interstellar medium (ISM) by stellar winds and PN ejection (Busso et al.\ 1999; Busso 2006, these proceedings).

Abundance determinations in PNe can be used to investigate TDU and AGB nucleosynthesis in their progenitors.  However, the difficulty of accurately determining C abundances in ionized nebulae (e.g.\ Kaler 1983) and the weakness of \emph{n}-capture element emission lines (due to their low cosmic abundances) have inhibited such studies in the past.  It was not until 1994 that \emph{n}-capture elements were identified in the spectrum of a PN (P\'{e}quignot \& Baluteau 1994).  This led Dinerstein (2001) to the realization that two anonymous features seen in the $K$ band spectra of PNe are in fact fine-structure transitions of the \emph{n}-capture elements Kr ($Z$=36) and Se ($Z$=34).

We have conducted a large-scale survey of [Kr~III]~2.199 and [Se~IV]~2.287~$\mu$m in Galactic disk PNe.  Because Kr and Se are not depleted into dust (Kr is a noble gas, and Se is not depleted in the ISM; Cardelli et al.\ 1993), they are useful tracers of \emph{s}-process enrichments and TDU in PN progenitors.  In spite of their small cosmic abundances ($\sim2\times10^{-9}$ relative to H), we have found that Se and Kr are detectable in a large fraction of PNe, and our survey has increased the number of PNe with known \emph{n}-capture element abundances by more than a factor of ten.

\vspace{-0.1in}
\section{Observations}

Using the CoolSpec spectrometer (Lester et al.\ 2000) on the 2.7-m telescope at McDonald Observatory, we have observed $\sim$100 PNe from 2.14--2.30~$\mu$m.  We observed all objects with a 2.7'' slit at a resolution $R=500$, which is sufficient to resolve the Kr and Se lines from other nearby features, except in PNe exhibiting vibrationally-excited H$_2$ emission.  We re-observed these objects with a high-resolution (1.0'' slit, $R=4400$) setting from 2.155--2.205~$\mu$m, allowing us to resolve [Kr~III] from H$_2$~3--2$S$(3)~2.201~$\mu$m.  We estimated the contribution of H$_2$~3--2$S$(2) to the flux of the 2.287~$\mu$m feature from the strength of H$_2$~3--2$S$(3) or other detected H$_2$ lines in the spectra.

We expanded our sample with $K$ band spectra from the literature for $\sim$20 additional PNe (Geballe et al.\ 1991; Hora et al.\ 1999; Lumsden et al.\ 2001).  We therefore have near-infrared spectra of 120 PNe, in which we have detected Se and/or Kr in 79.

\vspace{-0.1in}
\section{Abundance Determinations and Correlations}

We have derived Kr$^{++}$/H$^+$ and Se$^{3+}$/H$^+$ ionic abundances or upper limits for all objects in our sample.  A five-level model atom was used for Kr$^{++}$ and a two-level system for Se$^{3+}$, with transition probabilities from Bi\'{e}mont \& Hansen (1986; 1987) and the collision strengths of Sch\"{o}ning (1997) and Butler (2006, in preparation).

Determining the elemental abundances requires correcting for the abundances of unseen ionization stages.  We have determined ionization corrections for Se and Kr using the photoionization code CLOUDY (Ferland et al.\ 1998), which we modified by adding Se and Kr to the atomic database.  However, this method is complicated by the lack of atomic data governing the ionization balance of Se and Kr.  To reduce the uncertainty in the Kr ionization correction, we have modeled in detail 10 PNe with deep optical spectra exhibiting [Kr~IV] emission.  We empirically adjusted the approximate Kr atomic data so that the models reproduce the observed [Kr~III] and [Kr~IV] emission lines in these 10 objects.  However, no optical Se lines have been identified, and therefore our derived Se abundances are less accurate than those of Kr.  We then constructed a grid of CLOUDY models in order to derive ionization corrections.  These results are more robust than our previous abundance determinations (e.g.\ Sterling \& Dinerstein 2005), where we assumed that the Kr$^{++}$ and Se$^{3+}$ ionic fractions are the same as ions of other elements with similar ionization potential ranges.  The details of our photoionization modeling will be discussed in a forthcoming paper (Sterling et al.\ 2006, in preparation).

Using these results, we have determined Se and Kr abundances in each observed PN.  Se and Kr are enriched in 73\% of the PNe in which they were detected, and display a range of abundances from slightly subsolar to enhanced by a factor of 10 or more.  These enrichments indicate that the \emph{s}-process and TDU occurred in the progenitor stars.

In the 20 PNe displaying emission from both [Kr~III] and [Se~IV], we find [Kr/Se]~=~0.33.  This is in excellent agreement with theoretical \emph{s}-process models (Busso et al.\ 2001; Gallino 2005, priv.\ comm.) for the metallicity range of the observed PNe ($\sim$0.3--1.0~$Z_{\odot}$).  These models predict [Kr/Se]~=~0.0--0.5, depending on the mass of the layer experiencing \emph{s}-process nucleosynthesis and the initial progenitor mass.

We have searched for correlations between the Se and Kr abundances and other nebular properties.  In particular, Se and Kr are strongly enriched in Type~I PNe, which are believed to descend from more massive progenitors (M$\geq2.5$--4.0~M$_{\odot}$; Peimbert 1978), relative to other PNe (Table~1).  While the short interpulse times and small intershell masses diminish the role of the $^{13}$C neutron source in such intermediate-mass stars (Goriely \& Siess 2005; Busso 2006, priv.\ comm.), their high intershell temperatures activate the $^{22}$Ne neutron source.  This produces \emph{s}-process enrichments significantly different from the $^{13}$C source, with light \emph{n}-capture elements (such as Se and Kr) much more enriched than heavier species.  This correlation provides some of the first evidence (see also Garc\'{i}a-Lario et al., these proceedings) that the $^{22}$Ne neutron source is active in intermediate-mass AGB stars, which may be important sources of light \emph{n}-capture elements in the Universe.

\begin{table}[t]
\begin{center}
\caption{Se and Kr Abundances Vs.\ Nebular Properties}
\begin{tabular}{lcccc}
\hline
\multicolumn{1}{l}{} & \multicolumn{1}{c}{Derived} & \multicolumn{1}{c}{Number of} & \multicolumn{1}{c}{Derived} & \multicolumn{1}{c}{Number of} \\
\multicolumn{1}{l}{Property} & \multicolumn{1}{c}{$<$Se/H$>$} & \multicolumn{1}{c}{Se Detections} & \multicolumn{1}{c}{$<$Kr/H$>$} & \multicolumn{1}{c}{Kr Detections} \\
\hline
$[$WC$]$ & 4.7$\pm$0.7 (-9) & 16 & 2.3$\pm$0.8 (-8) & 8\\
Non-[WC] & 5.3$\pm$0.6 (-9) & 39 & 1.3$\pm$0.2 (-8) & 17\\
\hline
Type I & 5.5$\pm$0.7 (-9) & 11 & 2.5$\pm$1.0 (-8) & 5\\
Non-Type I & 4.8$\pm$0.4 (-9) & 56 & 1.3$\pm$0.2 (-8) & 22\\
\hline
O-rich Dust & 1.0$\pm$0.6 (-9) & 2 & ... & 0 \\
C-rich Dust & 4.2$\pm$0.7 (-9) & 13 & 1.5$\pm$0.4 (-8) & 13 \\
Mixed Dust & 2.4$\pm$1.2 (-9) & 3 & 8.1$\pm$1.8 (-9) & 5 \\
\hline
Full Sample & 4.9$\pm$0.4 (-9) & 67 & 1.6$\pm$0.3 (-8) & 27\\
Solar$^{a}$ & 2.14$\pm$0.20 (-9) & ... & 1.91$\pm$0.36 (-9) & ...\\
\hline
\end{tabular}
\end{center}
\hspace{0.5in}$^{a}$ From Asplund et al.\ (2005)
\vspace{-0.1in}
\end{table}

We also find that PNe with H-deficient Wolf-Rayet-like ([WC]) central stars tend to be more enriched in Kr than objects with H-rich nuclei (Table~1).  However, such a correlation is not seen in Se.  The reason for this is not clear; possibilities include the larger uncertainties in our Se abundance determinations and the potential difficulty of discerning correlations with Se, which is less enriched than Kr.  Regardless, [WC] PNe deserve further investigation to determine whether they are indeed highly \emph{s}-process enriched.

The Se and Kr abundances are expected to correlate with the C/O ratio, since TDU conveys carbon along with \emph{s}-processed material to the surface of AGB stars.  We plot [Kr/H] and [Se/H] vs.\ C/O in Figure~1.  Despite the considerable scatter, there appears to be a positive correlation between the \emph{n}-capture element abundances and C/O.  The relation between Se and Kr enrichments with C/O is strengthened by correlations with the dust composition (Table~1).  In particular, Se is more enriched in PNe with C-rich dust than those with O-rich or mixed (C- and O-rich) dust types.  Similarly, Kr is more enriched in PNe with C-rich dust than those with mixed dust chemistries.

\begin{figure}
\scalebox{0.8}{\includegraphics{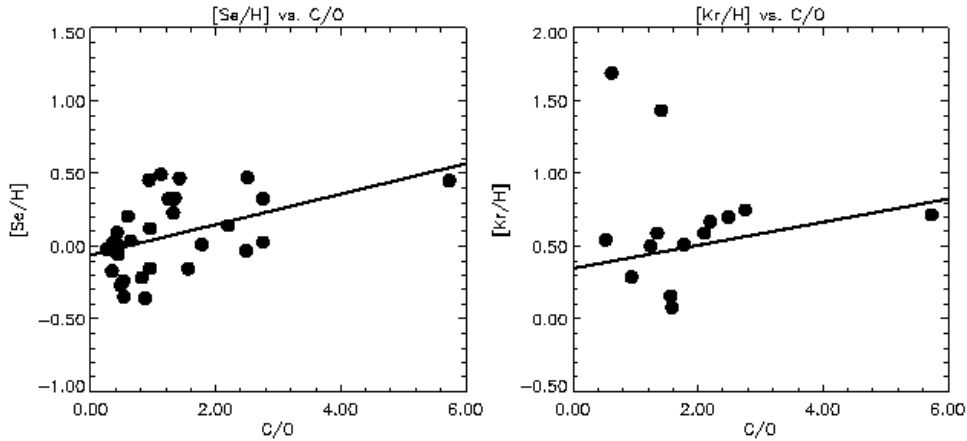}}
\caption{Se and Kr abundances vs.\ C/O.  The solid lines are linear fits (excluding the two outliers, both Type~I PNe, with high [Kr/H] and low C/O).}
\end{figure}

\vspace{-0.2in}
\section{Summary}

We have presented preliminary results from the first large-scale survey of \emph{n}-capture elements in PNe.  We have detected near-IR emission lines from Se and/or Kr in nearly two-thirds of our sample of 120 PNe.  Se and Kr are enriched in 73\% of the objects in which they were detected, which we attribute to the occurrence of \emph{s}-process nucleosynthesis and TDU in their progenitor stars.  Se and Kr tend to be more enriched in Type~I PNe than in other objects, which may be evidence for the increasing role of the $^{22}$Ne neutron source in intermediate-mass PN progenitors.  We also find that Se and Kr are positively correlated with C/O, as expected from the relation of the \emph{s}-process with TDU.  Our study is the first broad characterization of \emph{s}-process enrichments in PNe as a population, and reveals the history of nucleosynthesis and dredge-up during the AGB phase of PN progenitors.

\begin{acknowledgments}

This work has been supported by NSF grant AST 04-06809.

\end{acknowledgments}

\end{document}